%% file: Throughput-Analysis-v14.tex
\documentclass[journal,twocolumn,10pt]{IEEEtran}
\usepackage{graphicx}
\usepackage{amssymb}
\usepackage{amsmath}
\usepackage{mathtools}
\usepackage{dsfont}
\usepackage{cite}
\usepackage{stfloats}
\usepackage{subfigure}
\usepackage{psfrag}
\usepackage[mathscr]{euscript}
\usepackage{acronym}  
\usepackage{booktabs}
\usepackage{bbm}

\usepackage{float}
\usepackage{balance}
\input{SupportDocuments/aconym}
\input{SupportDocuments/defmetric}

\newcommand{\paperTitle}{Packet Throughput Analysis of Static and\\Dynamic TDD in Small Cell Networks}

\begin{document}

{
\title{\paperTitle}

\author{

	    Howard~H.~Yang,
        Giovanni Geraci,
        Yi~Zhong,
        and Tony~Q.~S.~Quek
\thanks{H.~H.~Yang and T.~Q.~S.~Quek are with the Information Systems Technology and Design Pillar, Singapore University of Technology and Design, Singapore (e-mail: hao\_yang@mymail.sutd.edu.sg, tonyquek@sutd.edu.sg).}
\thanks{G.~Geraci is with the Department of Small Cells Research, Nokia Bell Labs, Dublin, Republic of Ireland (e-mail: dr.giovanni.geraci@gmail.com).}
\thanks{Y.~Zhong is with the School of Electronic Information and Communications, Huazhong University of Science and Technology, Wuhan, P.R. China (email: yzhong@hust.edu.cn).}}
\maketitle
\acresetall
\thispagestyle{empty}
\begin{abstract}
We develop an analytical framework for the performance comparison of small cell networks operating under static time division duplexing (S-TDD) and dynamic TDD (D-TDD). While in S-TDD downlink/uplink (DL/UL) cell transmissions are synchronized, in D-TDD each cell dynamically allocates resources to the most demanding direction. By leveraging stochastic geometry and queuing theory, we derive closed-form expressions for the UL and DL packet throughput, also capturing the impact of random traffic arrivals and packet retransmissions. Through our analysis, which is validated via simulations, we confirm that D-TDD outperforms S-TDD in DL, with the vice versa occurring in UL, since asymmetric transmissions reduce DL interference at the expense of an increased UL interference. We also find that in asymmetric scenarios, where most of the traffic is in DL, D-TDD provides a DL packet throughput gain by better controlling the queuing delay, and that such gain vanishes in the light-traffic regime.
\end{abstract}
\begin{IEEEkeywords}
Dynamic time division duplexing, small cells, packet throughput, stochastic geometry, queuing theory.
\end{IEEEkeywords}

\acresetall

\section{Introduction}\label{sec:intro} 

As well as continuously increasing, the wireless data demand is shifting from symmetric downlink/uplink (DL/UL) capacity requirements, e.g., for voice traffic, to strongly asymmetric and fluctuating activities, e.g., video streaming or file uploading \cite{Nomor2015}. It is becoming crucial to allocate spectrum resources between the UL and the DL based on immediate traffic needs, and new features are being incorporated into Long Term Evolution (LTE) to allow for a more flexible use of radio resources, such as the enhanced Interference Mitigation and Traffic Adaptation (eIMTA) \cite{3GPP36300R12}. This flexible DL/UL capacity split, commonly referred to as \emph{dynamic time division duplexing} (D-TDD), is also expected to be one of the operation modes for fifth-generation (5G) ultra-dense networks \cite{LahPajVih2014,Nokia2015}.

Unlike conventional \emph{static TDD} (S-TDD), which requires all DL/UL cell activities to be synchronized, D-TDD allows each cell to individually configure its subframe to accomodate whichever link direction needs it the most \cite{SheKhoEri:12}. As a result, D-TDD may provide higher spectrum utilization and reduced latency, and it is particularly appealing for network scenarios with significant traffic fluctuation. On the other hand, D-TDD suffers additional inter-cell interference introduced by asynchronous UL/DL transmissions, and it may not be suitable for all small cell deployment configurations \cite{LopDinCla:15}.

System-level comparisons between S-TDD and D-TDD have been performed, among others, in terms of coverage probability \cite{ZhoCheWan:15}, achievable rate \cite{GupKulVis:16}, and energy efficiency \cite{SunSheWil:16}, showing the gains attainable by D-TDD. In particular, significant improvements have been demonstrated in the presence of interference mitigation techniques, e.g., power control, cell clustering, or interference cancellation \cite{DinPorVas:14}.
While these previous works provide a basic assessment of the performance of D-TDD vs. S-TDD, the full-buffer assumption commonly used fails to capture the crucial effect of queuing delay, in turn affected by random packet arrivals and retransmissions \cite{ZhoQueGe:16}.

In this paper, we overcome such limitation by adopting the mean packet throughput as the performance metric -- capturing the effect of both transmission and queuing delay -- and we propose a stochastic geometry framework that quantifies the impact of various network parameters. In particular, we model the locations of \acp{SAP} and user equipment (UEs) as independent \acp{PPP}, the UL/DL traffic arrivals as independent Bernoulli processes \cite{ZhoQueGe:16}, and account for the retransmission of unsuccessfully delivered packets. We derive accurate closed-form expressions for the mean packet throughput under S-TDD and D-TDD, allowing to compare them and to draw insightful conclusions.

\section{System Model}\label{sec:sysmod}

\subsection{Network Topology and Scheduling}

We consider a small cell network that consists of \acp{SAP} and UEs. The spatial locations of  SAPs and UEs follow independent \acp{PPP} $\Phi_{\mathrm{s}}$  and $\Phi_{\mathrm{u}}$, with spatial densities  $\LS$ and $\LU$, respectively.
All  SAPs and UEs are equipped with a single antenna, and transmit with power $\PST$ and $\PUT$, respectively.
The channels between any pair of nodes are modeled as independent and identically distributed (i.i.d.) and quasi-static. We assume each channel to be narrowband and affected by two attenuation components, namely small scale Rayleigh fading, and large-scale path loss.\footnote{The results obtained in this paper through the machinery of stochastic geometry can be extended to account for the presence of shadowing \cite{YanGerQue:16}.} Furthermore, UEs associate to the SAPs that provides the largest average received power. Since the association policy can result in multiple UEs associating to one \ac{SAP}, we limit the maximum number of UEs served by each \ac{SAP} (denoted by $N_{\mathrm{s}}$) to $K_{\mathrm{s}}$, and assume that each SAP randomly select one of its served UEs at each time slot.

\subsection{Traffic Model}

We use a discrete time queuing system to model the traffic profile. In particular, the time axis is segmented into a sequence of equal intervals, referred to as time slots. We assume that all queuing activities, i.e., packet arrivals and departures, take place around the slot boundaries. Specifically, at the $n$-th time slot, a potential packet departure may occur in the interval $(n^{-},n)$, and a potential packet arrival may happen in the interval $(n,n^+)$. In other words, departures occur immediately before slot boundaries, whereas arrivals occur immediately after slot boundaries.

For a generic UE, we model its UL/DL packet arrivals as independent Bernoulli processes with rates $\xi_{\mathrm{U}}, \xi_{\mathrm{D}} \in [0,1]$, respectively, representing the probability an arrival occurs in a time slot \cite{StaHae:10}. Moreover, we assume that each node accumulates all incoming packets in an infinite-size buffer. 

\subsection{Radio Access}

We consider two TDD modes of operation for radio access, i.e., S-TDD and D-TDD, described as follows \cite{SheKhoEri:12}.

\subsubsection{S-TDD} At each time slot, all \acp{SAP} transmit either in DL or in UL with probabilities $p_\mathrm{S}$ and $1-p_\mathrm{S}$, respectively.
\subsubsection{D-TDD} \acp{SAP} independently schedule their transmissions. In a given time slot, a typical \ac{SAP} transmits in DL (resp. UL) with probability $p_{\mathrm{D}}$ (resp. $1- p_{\mathrm{D}}$).

\section{Analysis}

\subsection{Preliminaries}

\subsubsection{Signal-to-interference ratio (SIR)}
Let $\zeta_{x,t} \in \{0,1\}$ be an indicator showing whether a node located at $x \in \Phi \triangleq \Phi_{\mathrm{s}} \cup \Phi_{\mathrm{u}}$ is transmitting at time slot $t$ ($\zeta_{x,t} = 1$) or not ($\zeta_{x,t}=0$). By the Slivyark's theorem, we can focus on a typical UE located at the origin and served by BS $x_0$. In an interference-limited network, the effect of thermal noise can be neglected, and the received DL {SIR} under S-TDD and D-TDD can be respectively written as
\begin{align}
\gamma_{\mathrm{S}, t}^{\mathrm{D}} &= \frac{\PST h_{x_0} \Vert x_0 \Vert^{-\alpha} }{  \sum\limits_{x \in \Phi_{\mathrm{s}}\setminus x_0  } \!\!\!\! \frac{\PST \zeta_{x,t} h_{x}}{ \Vert x \Vert^{\alpha}}   }, \label{eqn:gammaSD}\\
\gamma_{\mathrm{D}, t}^{\mathrm{D}} &= \frac{\PST h_{x_0} \Vert x_0 \Vert^{-\alpha} }{  \sum\limits_{x \in \Phi_{\mathrm{s}}\setminus x_0  } \!\!\!\! \frac{\PST \zeta_{x,t} h_{x}}{ \Vert x \Vert^{\alpha}}  +\!\! \sum\limits_{z \in \Phi_{\mathrm{u}} } \!\!\!\! \frac{ \PUT \zeta_{z,t} h_{z}}{ \Vert z \Vert^{\alpha}}   }. \label{eqn:gammaDD}
\end{align}
Similarly, the UL {SIR} under S-TDD and D-TDD received by a typical \ac{SAP} from UE $z_0$ can be respectively expressed as
\begin{align}
\gamma_{\mathrm{S}, t}^{\mathrm{U}} &= \frac{\PUT h_{z_0} \Vert z_0 \Vert^{-\alpha} }{  \sum\limits_{z \in \Phi_{\mathrm{u}}\setminus z_0  } \!\!\!\! \frac{\PST \zeta_{z,t} h_{z}}{ \Vert z \Vert^{\alpha}}   },\label{eqn:gammaSU}\\
\gamma_{\mathrm{D}, t}^{\mathrm{U}} &= \frac{\PUT h_{z_0} \Vert z_0 \Vert^{-\alpha} }{  \sum\limits_{z \in \Phi_{\mathrm{u}} \setminus z_0 } \!\!\!\! \frac{ \PUT \zeta_{z,t} h_{z}}{ \Vert z \Vert^{\alpha}}  + \sum\limits_{x \in \Phi_{\mathrm{s}} } \!\!\!\! \frac{\PST \zeta_{x,t} h_{x}}{ \Vert x \Vert^{\alpha}}    }.\label{eqn:gammaDU}
\end{align}

\subsubsection{Transmission and Success Probability}

During each time slot, every node with a non-empty buffer sends out a packet from the head of its queue. If the received SIR exceeds a predefined threshold, the transmission is successful and the packet can be removed from the queue; otherwise, the transmission fails and the packet remains in the buffer. The success probability $\mu_{t}$ is therefore defined as the probability that the received {SIR} $\gamma_t$ is above a certain threshold $\theta$, i.e.,
\begin{align}
\mu_{t} = \mathbb{P}\left(\gamma_{t} > \theta \right).
\end{align}
We note that the success probability can be regarded equivalently as the service rate of the queuing system.

\subsubsection{Mean Packet Throughput}

We employ packet throughput, i.e., the number of successfully transmitted packets per time slot, as our performance metric. A formal definition is given as follows. 

\begin{definition}
\textit{
Let $A_x(t)$ be the number of packets arrived at a typical transmitter $x$ within period $[0,t]$, and $D_{i, x}$ be the number of time slots between the arrival of the $i$-th packet and its successful delivery. The mean packet throughput is defined as
\begin{align}\label{equ:Gen_PktThrPut}
\mathcal{T} \triangleq \lim_{R \rightarrow \infty} \frac{\sum_{x \in \Phi \cap B(0,R)}  \lim\limits_{t \rightarrow \infty} \frac{A_x(t)}{\sum_{i=1}^{A_x(t)} D_{i, x} }}{\sum_{x \in \Phi} \mathbbm{1}_{\{ x \in B( 0, R ) \}} }.
\end{align}
}
\end{definition}
Note that $D_{i,x}$ in \eqref{equ:Gen_PktThrPut} represents the number of time slots required to successfully deliver the $i$-th packet, and its value is affected by: $(i)$ queueing delay, caused by other accumulated unsent packets, and $(ii)$ transmission delay, due to link failure and retransmission.
By averaging over all nodes, \eqref{equ:Gen_PktThrPut} provides information on the packet throughput across the network.

\subsection{Packet Throughput Analysis}

This section details the main results of our work. We first introduce two lemmas to facilitate the analysis. 
\begin{lemma}\label{prop:Con_RS}
\textit{
Given the number of served UEs $N_{\mathrm{s}}$, the arrival rate $\xi_x$, and the service rate $\mu$, the mean packet throughput at a typical SAP is
\begin{align}\label{equ:CndAveThrPut}
\mathcal{T} = \left\{ \frac{\mu/N_{\mathrm{s}} -  \xi_x}{1 - \xi_x} \right\}_{+}, ~~~x \in \{\mathrm{U}, \mathrm{D}\}, 
\end{align}
and its idle probability is given by
\begin{align}
\tau_0 = \left\{1 - {{N_{\mathrm{s}}} \xi_x}/{\mu} \right\}_{+},
\end{align}
where $\{x\}_{+} \triangleq \max\{x,0\}$.
}
\end{lemma}
\begin{IEEEproof}
See Appendix~\ref{apx:Con_RS} for a sketch of the proof.
\end{IEEEproof}
\begin{lemma}\label{lma:UE_Num}
\textit{
The probability mass function (PMF) of the number of served UEs per SAP, $N_{\mathrm{s}}$, is given by 
\begin{align}\label{equ:UE_Num}
f_{N_{\mathrm{s}}}(i) = 
\left\{
       \begin{array}{ll}
          \frac{\eta^\eta \Gamma\left( i + \eta \right) \rho^{-i} }{i! \Gamma(\eta) \left( \frac{1}{\rho} + \eta \right)^{i+\eta} } \quad \quad \qquad~ \text{if}~~ i \leq K_{\mathrm{s}} - 1, \\
          \sum_{j=K_{\mathrm{s}}}^{\infty} \frac{\eta^\eta \Gamma\left( j + \eta \right) \rho^{-j} }{j! \Gamma(\eta) \left( \frac{1}{\rho} + \eta \right)^{j+\eta} } \quad \text{if} ~ i = K_{\mathrm{s}}
       \end{array}
\right.
\end{align}
where $\rho = \LS/\LU$, $\eta = 3.5$, and $\Gamma(\cdot)$ is the Gamma function. }
\end{lemma}
\begin{IEEEproof}
See \cite{YanGerQue:16} for a detailed proof.
\end{IEEEproof}

The average packet throughput under S-TDD can then be derived as follows. 
\begin{theorem}\label{thm:Pkt_ThrPut}
\textit{
The mean UL and DL packet throughput under S-TDD can be respectively approximated as
\begin{align}
\mathcal{T}_{\mathrm{S}}^{\mathrm{U}}&\!\approx\! \sum_{i=1}^{\KS} \frac{f_{N_{\mathrm{s}}}(i)}{{1 \!-\! \xi_{\mathrm{U}} } } \! \left\{ { \frac{1\!-\!p_\mathrm{S}}{i} - \xi_{\mathrm{U}}\! \left(\! 1 \!+\! \frac{\mathbb{E}\!\left[ N_{\mathrm{s}} \right] \mathcal{V}\!\left( \theta, \alpha \!\right)}{i}  \right) }\! \right\}_{\!+}\!, \\
\mathcal{T}_{\mathrm{S}}^{\mathrm{D}}&\!\approx\! \sum_{i=1}^{\KS} \frac{f_{N_{\mathrm{s}}}(i)}{ {1 \!-\! \xi_{\mathrm{D}} } } \! \left\{ { \frac{p_\mathrm{S}}{i} - \xi_{\mathrm{D}}\! \left( 1 \!+\! \frac{\mathbb{E}\!\left[ N_{\mathrm{s}} \right] \mathcal{Z}\!\left( \theta, \alpha \right)}{i}  \right) }\! \right\}_{\!+}
\end{align}
where $\mathcal{V}(\theta, \alpha)$ and $\mathcal{Z}\left( \theta, \alpha \right)$ are given as follows
\begin{align}
\mathcal{V}\left( \theta, \alpha \right) = \frac{ 2 \pi  \theta^{\frac{2}{\alpha}} }{ \alpha \sin(\frac{2 \pi }{\alpha})}, ~~
\mathcal{Z}\left( \theta, \alpha \right) =\! \int_{\theta^{-\frac{2}{\alpha}}}^{\infty} \frac{\theta^{\frac{2}{\alpha}} du }{1+u^{\frac{\alpha}{2}}}.
\end{align}
}
\end{theorem}
\begin{IEEEproof}
See Appendix~\ref{apx:Pkt_ThrPut} for a sketch of the proof.
\end{IEEEproof}

In regard to D-TDD, we assume that each cell individually allocates its DL time fraction $p_\mathrm{D}$ to minimize the average DL/UL traffic demand \cite{DinPorVas:14}, i.e.,
\begin{align} \label{equ:DL_Prt}
p_{\mathrm{D}} = \arg\min_{p \in [0,1]} \left| \frac{\xi_{\mathrm{D}} }{p} - \frac{\xi_{\mathrm{U}} }{1-p} \right|.
\end{align}
Solving \eqref{equ:DL_Prt} yields $p_{\mathrm{D}} = \xi_{\mathrm{D}}/(\xi_{\mathrm{U}} + \xi_{\mathrm{D}} )$ for all SAPs, and the mean D-TDD packet throughput can be derived as follows. 
\begin{theorem}\label{thm:Pkt_ThrPut_DTDD}
\textit{
The mean UL and DL packet throughput under D-TDD can be respectively approximated as
\begin{align}
\mathcal{T}_{\mathrm{D}}^{\mathrm{U}} &\approx \sum_{k=1}^{K_{\mathrm{s}}} \left( 1 - p_{\mathrm{D}} \right) \cdot f_{N_{\mathrm{s}}}(k) \cdot \left\{ \frac{ \frac{\mu_{\mathrm{U}}}{k} - \xi_{\mathrm{U}} }{1 - \xi_{\mathrm{U}} } \right\}_{+},\\
\mathcal{T}_{\mathrm{D}}^{\mathrm{D}} &\approx \sum_{k=1}^{K_{\mathrm{s}}} p_{\mathrm{D}} \cdot f_{N_{\mathrm{s}}}(k) \cdot \left\{ \frac{ \frac{\mu_{\mathrm{D}}}{k} - \xi_{\mathrm{D}} }{1 - \xi_{\mathrm{D}} } \right\}_{+},
\end{align}
with $p_{\mathrm{D}}$ as in (\ref{equ:DL_Prt}), and where $\mu_{\mathrm{D}}$ and $\mu_{\mathrm{U}}$ are given as follows
\begin{align}
\mu_{\mathrm{U}} &= \frac{ \xi_{\mathrm{U}} \!+\! \xi_{\mathrm{D}} - \left[ \xi_{\mathrm{D}}^2 \mathcal{Z}(\theta, \alpha) + \xi_{\mathrm{U}}^2 \mathcal{V}(\theta, \alpha) \right]\! \mathbb{E}[N_{\mathrm{s}}] }{ \xi_{\mathrm{U}} \!+\! \xi_{\mathrm{D}} - \xi_{\mathrm{D}}^2 \mathbb{E}[N_{\mathrm{s}}]\! \left[ \mathcal{Z}(\theta, \alpha) - \mathcal{V}(\theta, \alpha) \frac{P_{\mathrm{st}}}{ P_{\mathrm{ut}} } \right] } ,\\
\mu_{\mathrm{D}} &= \frac{ \xi_{\mathrm{U}} \!+\! \xi_{\mathrm{D}} - \left[ \xi_{\mathrm{D}}^2 \mathcal{Z}(\theta, \alpha) + \xi_{\mathrm{U}}^2 \mathcal{V}(\theta, \alpha) \right]\! \mathbb{E}[N_{\mathrm{s}}]  }{\xi_{\mathrm{D}} \!+\! \xi_{\mathrm{U}} - \xi_{\mathrm{U}}^2 \mathbb{E}\!\left[ N_{\mathrm{s}} \right] \mathcal{V}\!\left( \theta, \alpha \right)\! \left( 1 - \frac{P_{\mathrm{ut}}}{P_{\mathrm{st}}} \right) }.
\end{align}
}
\end{theorem}
\begin{IEEEproof}
See Appendix~\ref{apx:Pkt_ThrPut_DTDD} for a sketch of the proof.
\end{IEEEproof}

\section{Numerical Results}

Unless otherwise stated, we adopt the following system parameters \cite{YanGerQue:16}: $\LS = 10^{-4} \mathrm{m}^{-2}$, $\LU = 10^{-3} \mathrm{m}^{-2}$, $\PST = 23$~dBm, $\PUT=17$~dBm, $K_{\mathrm{s}}=3$, $\theta = 0$~dB, and $\alpha = 3.8$. Moreover, we set the DL time portion for both S-TDD and D-TDD to be the same, i.e., $p_{\mathrm{S}}=p_{\mathrm{D}}$, and as per \eqref{equ:DL_Prt}. 

\begin{figure}[t!]
  \centering{}

    {\includegraphics[width=1.0\columnwidth]{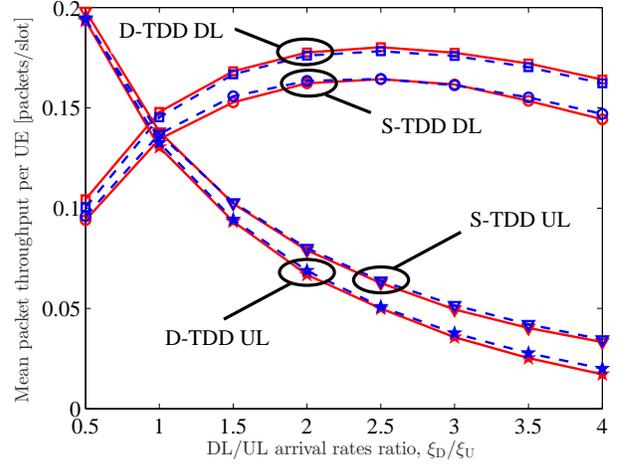}}

  \caption{Mean packet throughput per UE vs. DL/UL arrival rates ratio: analytical results (dashed lines) and simulations (solid lines).}
  \label{fig:TDDPktThrPut}
\end{figure}

In Fig.~\ref{fig:TDDPktThrPut}, we depict the mean throughput per UE, expressed in packets per time slot. In this figure, the UL arrival rate is kept constant as $\xi_{\mathrm{U}} = 0.02$, and the DL arrival rate is varied to show its effect. The figure shows that analytical results (dashed lines) and simulations (solid lines) well match, validating Theorem~1 and Theorem~2. Several observations are due:
$(i)$ as the DL arrival rate grows from low to medium values, the DL throughput increases since a larger portion of time slots is allocated to DL transmissions;
$(ii)$ as the DL arrival rate exceeds a certain value, many packets start accumulating in the buffer, and the queuing delay starts degrading the system performance, yielding a decreasing throughput;
$(iii)$ since the UL arrival rate is fixed, increasing the DL arrival rate causes the UL throughput to decrease; and
$(iv)$ consistently with \cite{DinPorVas:14}, D-TDD outperforms S-TDD in DL (especially when DL traffic is prevalent), and vice versa in UL, since asymmetric transmissions reduce DL interference at the expense of an increased UL interference.

\begin{figure}[t!]
  \centering{}

    {\includegraphics[width=1.0\columnwidth]{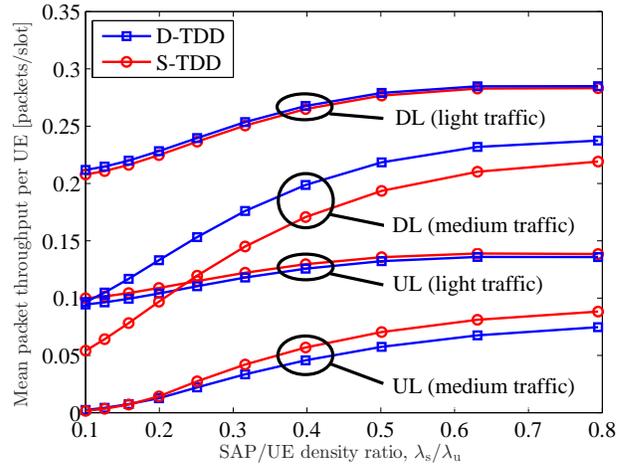}}

  \caption{Mean packet throughput per UE versus SAP/UE density ratio.}
  \label{fig:Densification}
\end{figure}

Fig.~\ref{fig:Densification} illustrates the mean packet throughput per UE, obtained via Theorems 1 and 2, for two different traffic statistics, namely, ($\xi_{\mathrm{U}} = 0.005, \xi_{\mathrm{D}}=0.01$) and ($\xi_{\mathrm{U}} = 0.05, \xi_{\mathrm{D}}=0.1$), respectively labeled as light and medium traffic. In this figure, the UE density is kept constant as $\LU = 10^{-3} \mathrm{m}^{-2}$ and the SAP density is varied to show its effect. Fig.~\ref{fig:Densification} shows that, although both S-TDD and D-TDD benefit from cell densification, their relative performance varies with the traffic conditions, namely:
$(i)$ in the presence of light traffic, i.e., small packet arrival rates, S-TDD and D-TDD exhibit very similar throughput, since nodes have short queues, and all packets can be transmitted quickly irrespective of the TDD mode; and
$(ii)$ in a medium-traffic regime, the effect of the TDD mode is more pronounced, especially in DL where D-TDD outperforms S-TDD by adaptively following the traffic fluctuations, and thus controlling the queuing delay.


\section{Conclusion}

We evaluated the performance of small cell deployments under static and dynamic TDD. For networks where topology, traffic arrivals, and scheduling are random, we analyzed the packet throughput also accounting for retransmissions and queuing delay. We confirmed that S-TDD outperforms D-TDD in the UL, while the opposite is true for DL operations. We also showed that the gain provided by D-TDD in asymmetric, DL-dominated scenarios, is more significant under moderate traffic, whereas it vanishes in the light-traffic regime.

While our emphasis was on the impact of the traffic pattern and queuing delay on the packet throughput, we note that the latter is also affected by the propagation environment and by the deployment scenario. Analyzing the performance of both TDD modes in networks with line-of-sight (LoS)-non-LoS transitions, and under a more realistic Poisson cluster process (PCP) model is regarded as an important research direction.

\begin{appendix}

\subsection{Sketch of Proof of Lemma~\ref{prop:Con_RS}}\label{apx:Con_RS}
A generic queue in the system is equivalent to a Geo/G/1 queue with arrival $\xi_x$ and service rate $\mu/N_{\mathrm{s}}$. Denoting $\pi_x(j)$ the probability of having $j$ packets in the steady state, the balance equation can be written as
\begin{align}
&\pi_x(0) \xi_x = \pi_x(1) (1-\xi_x) \frac{\mu}{N_{\mathrm{s}}} ~\Rightarrow ~ \pi_x(1) = \frac{N \xi_x}{1-\xi_x} \pi_x(0)
\nonumber\\
&\vdots
\nonumber\\
&\pi_x(j) = \frac{N_{\mathrm{s}} \xi_x^j \left( N_{\mathrm{s}} - \mu \right)^{j-1} }{ (1-\xi_x)^j \mu^j } \pi_x(0).
\end{align} 
We note that $\pi_x(j)$ can be obtained from $\sum_{j=0}^\infty \pi_x(j) = 1$. As such, the mean packet throughput can be derived as 
\begin{align}
\mathcal{T}_{\Phi} = \frac{\xi_{x}}{\sum_{j=0}^{\infty} \pi_x(j) \cdot j },
\end{align}
and the system idle probability is given by $\tau_0 = \pi_x(0)$.

\subsection{Sketch of Proof of Theorem~\ref{thm:Pkt_ThrPut}} \label{apx:Pkt_ThrPut}
Let us decompose the PPP $\Phi_{\mathrm{s}}$ into $\KS$ tiers, where the $k$-th tier consists of SAPs with $k$ associated UEs. As such, the location of $k$-th tier SAPs can be approximated by an independent PPP with spatial density $\LS f_{N_{\mathrm{s}}}(k)$. Next, the probability of a $k$-th tier SAP being active can be approximated as $\mathbb{P}(\zeta_{k} = 1) \approx \frac{k \xi_{\mathrm{D}} }{p_{\mathrm{S}} \mu_{\mathrm{S}}^{\mathrm{D}} }$. The approximated service rate at the typical UE is then obtained from (\ref{eqn:gammaSD}) as
\begin{align}
&\mu_{\mathrm{S}}^{\mathrm{D}} \approx \mathbb{E}\left[ \prod_{ k=1 }^{\KS} \exp\left( \sum_{x \in \Phi_{\mathrm{s}}^k \setminus \{x_0\} } \!\!\!\!\! - \frac{ \zeta_{x} h_x \theta \| x_0 \|^\alpha }{ \|x\|^\alpha } \right) \right]
\nonumber\\
&= \int_{0}^{\infty} \!\!\!\! \exp\!\left( \! -\LS \pi r^2 \!-\! \LS \pi r^2 \frac{ \xi_{\mathrm{D}} \sum_{k = 1}^{\KS} k f_{N_{\mathrm{s}}}(k) }{ p_{\mathrm{S}} \mu_{\mathrm{S}}^{\mathrm{D}} [\mathcal{Z}\!\left( \theta, \alpha \right)]^{-1}  }  \right) 2 \pi \LS r dr
\nonumber\\
&= \left( 1 + \frac{ \mathbb{E}\left[ N_{\mathrm{s}} \right] \xi_{\mathrm{D}} \mathcal{Z}\left( \theta, \alpha \right) }{ p_{\mathrm{S}} \mu_{\mathrm{S}}^{\mathrm{D}} } \right)^{-1}.
\end{align}
Solving the above equation and employing Lemma~\ref{prop:Con_RS} yields
\begin{align}
\mu_{\mathrm{S}}^{\mathrm{D}} \approx 1 - \frac{ \mathbb{E}\left[ N_{\mathrm{s}} \right] \xi_{\mathrm{D}} \mathcal{Z}\left( \theta, \alpha \right) }{p_{\mathrm{S}}},
\label{eqn:muSD}
\end{align}
\begin{align}
\mathcal{T}_{\mathrm{S}}^{\mathrm{D}} = \sum_{k=1}^{\KS} f_{N_{\mathrm{s}}}(k) \cdot \left\{ \frac{p_{\mathrm{S}} \mu_{\mathrm{S}}^{\mathrm{D}}/k - \xi_{\mathrm{D}} }{1-\xi_{\mathrm{D}}} \right\}_{+},
\end{align}
with $\mu_{\mathrm{S}}^{\mathrm{D}}$ as in (\ref{eqn:muSD}). The approximated UL packet throughput can be derived similarly from (\ref{eqn:gammaSU}).

\subsection{Sketch of Proof of Theorem~\ref{thm:Pkt_ThrPut_DTDD}} \label{apx:Pkt_ThrPut_DTDD}

Similarly to Appendix~\ref{apx:Pkt_ThrPut}, by leveraging (\ref{eqn:gammaDD}), (\ref{eqn:gammaDU}), and stochastic geometry tools, the service rate of DL/UL D-TDD transmissions can be respectively approximated as follows
\begin{align}\label{equ:DTDD_DL_SevRat}
\mu_{\mathrm{D}}^{\mathrm{D}} &= \mathbb{E}\!\! \left(\! \exp\!\! \left[ -{\theta \Vert x_0 \Vert^\alpha } \!\!\left( \sum_{x \in \Phi_{\mathrm{s}} \! \setminus \! \{ x_0 \} } \!\!\!\!\!\! \frac{ \zeta_{x,t} h_x }{ \Vert x \Vert^\alpha }  \!+\!\!\! \sum_{z \in \Phi_{\mathrm{u}}} \!\! \frac{P_{\mathrm{ut}} \zeta_{z,t} h_z }{ { P_{\mathrm{st}} } \Vert  z \Vert^\alpha } \! \right) \! \right] \! \right)
\nonumber\\
&\approx\!\! \left(\! 1 \!+\! \frac{p_{\mathrm{D}}  \xi_{\mathrm{D}} \mathcal{Z}(\theta, \alpha) }{ \mu_{\mathrm{D}}^{\mathrm{D}} \mathbb{E}[N_{\mathrm{s}}]^{-1} } \!+\! \frac{( 1 \!-\! p_{\mathrm{D}}) \xi_{\mathrm{U}} \mathcal{V}(\theta, \alpha) }{ \mu_{\mathrm{D}}^{\mathrm{U}} \mathbb{E}[N_{\mathrm{s}}]^{-1}  } \frac{P_{\mathrm{ut}}}{ P_{\mathrm{st}} } \right)^{-1}\!\!,\\ \label{equ:DTDD_UL_SevRat}
\mu_{\mathrm{D}}^{\mathrm{U}} \!&\approx \!\left(\! 1  \!+\! \frac{( 1 \!-\! p_{\mathrm{D}}) \xi_{\mathrm{D}} \mathcal{V}(\theta, \alpha) }{ \mu_{\mathrm{D}}^{\mathrm{D}} \mathbb{E}[N_{\mathrm{s}}]^{-1}  } \!+\! \frac{p_{\mathrm{D}}  \xi_{\mathrm{D}} \mathcal{V}(\theta, \alpha) }{ \mu_{\mathrm{D}}^{\mathrm{U}} \mathbb{E}[N_{\mathrm{s}}]^{\!-1} } \frac{P_{\mathrm{st}}}{ P_{\mathrm{ut}} } \right)^{-1}\!\!. 
\end{align}
The theorem then follows from solving the system of equations \eqref{equ:DTDD_DL_SevRat} and \eqref{equ:DTDD_UL_SevRat}, and applying  Lemma~\ref{prop:Con_RS}.

\end{appendix}

\bibliographystyle{IEEEtran}
\bibliography{bib/StringDefinitions,bib/IEEEabrv,bib/howard_trff}

\end{document}

%% file: SupportDocuments/aconym.tex
\acrodef{CCDF}{complementary cumulative distribution function}
\acrodef{CF}{characteristic function}
\acrodef{PPP}{Poisson point processe}
\acrodef{RV}{random variable}
\acrodef{i.i.d.}{independent and identically distributed}
\acrodef{PDF}{probability distribution function}
\acrodef{CDF}{cumulative distribution function}
\acrodef{ch.f.}{characteristic function}
\acrodef{AWGN}{additive white Gaussian noise}
\acrodef{SNR}{signal-to-noise ratio}
\acrodef{LRT}{likelihood ratio test}
\acrodef{DRT}{distance ratio test}
\acrodef{GLRT}{generalized likelihood ratio test}
\acrodef{CRLB}{Cram\'{e}r-Rao lower bound}
\acrodef{CRB}{Cram\'{e}r-Rao bound}
\acrodef{ZZLB}{Ziv-Zakai lower bound}
\acrodef{ZZB}{Ziv-Zakai bound}
\acrodef{LOS}{line-of-sight}
\acrodef{ToF}{time-of-flight}
\acrodef{NLOS}{non-line-of-sight}
\acrodef{GDOP}{geometric dilution of precision}
\acrodef{GPS}{Global Positioning System}
\acrodef{FIM}{Fisher information matrix}
\acrodef{PEB}{position error bound}
\acrodef{SPEB}{squared position error bound}
\acrodef{TOA}{time-of-arrival}
\acrodef{TOF}{time-of-flight}
\acrodef{WSN}{wireless sensor network}
\acrodef{MAC}{medium access control}
\acrodef{RSS}{received signal strength}
\acrodef{WAF}{wall attenuation factor}
\acrodef{TDOA}{time difference-of-arrival}
\acrodef{RF}{radiofrequency}
\acrodef{RTT}{round-trip time}
\acrodef{AOA}{angle-of-arrival}
\acrodef{MF}{matched filter}
\acrodef{ED}{energy detector}
\acrodef{ML}{maximum likelihood}
\acrodef{MSE}{mean-square error}
\acrodef{RMSE}{root-mean-square error}
\acrodef{LEO}{localization error outage}
\acrodef{ppm}{part-per-million}
\acrodef{ACK}{acknowledge}
\acrodef{UWB}{Ultrawide bandwidth}
\acrodef{TNR}{threshold-to-noise ratio}
\acrodef{LS}{least squares}
\acrodef{IR-UWB}{impulse radio UWB}
\acrodef{FCC}{Federal Communications Commission}
\acrodef{TH}{time-hopping}
\acrodef{PPM}{pulse position modulation}
\acrodef{MUI}{multi-user interference}
\acrodef{PDP}{power delay profile}
\acrodef{BPZF}{band-pass zonal filter}
\acrodef{SIR}{signal-to-interference ratio}
\acrodef{SINR}{signal-to-interference-plus-noise ratio}
\acrodef{RFID}{radio frequency identification}
\acrodef{WPAN}{wireless personal area network}
\acrodef{WWB}{Weiss-Weinstein bound}
\acrodef{DP}{direct path}
\acrodef{MF}{matched filter}
\acrodef{MMSE}{minimum-mean-square-error}
\acrodef{SBS}{serial backward search}
\acrodef{SBSMC}{serial backward search for multiple clusters}
\acrodef{NBI}{narrowband interference}
\acrodef{WBI}{wideband interference}
\acrodef{INR}{interference-to-noise ratio}
\acrodef{CR}{channel response}
\acrodef{CIR}{channel impulse response}
\acrodef{CR}{channel  response}
\acrodef{RADAR}{radar}
\acrodef{MUR}{Multistatic radar}
\acrodef{JBSF}{jump back and search forward}
\acrodef{HDSA}{high-definition situation-aware}
\acrodef{RRC}{root raised cosine}
\acrodef{ST}{simple thresholding}
\acrodef{BTB}{Bellini-Tartara bound}
\acrodef{P-Max}{$P$-Max}  
\acrodef{MIMO}{multiple-input multiple-output}
\acrodef{MAP}{maximum a posteriori}
\acrodef{FG}{factor graph}
\acrodef{OP}{outage probability}
\acrodef{WED}{wall extra delay}
\acrodef{RMS}{root mean square}
\acrodef{SPAWN}{sum-product algorithm over a wireless network}
\acrodef{MDD}{minimum distance distribution}
\acrodef{MAP}{maximum a posteriori probability}
\acrodef{SAP}{small cell access point}
\acrodef{UE}{user equipment}
\acrodef{MBS}{macro cell base station}
\acrodef{UER}{\ac{UE} Relay}
\acrodef{D2D}{device-to-device}
\acrodef{MBS}{macro base station}
\acrodef{CSI}{channel state information}
\acrodef{OGR}{outage guard region}
\acrodef{FUR}{feasible UER region}
\acrodef{EHR}{energy harvesting region}
\acrodef{EH}{energy harvesting}
\acrodef{D2D-EHSN}{D2D communication provided \ac{EH} small cell network}
\acrodef{D2D-EHHN}{D2D communication provided \ac{EH} heterogeneous network}
\acrodef{3GPP}{3rd Generation Partnership Project}
\acrodef{BS}{base station}
\acrodef{DF}{decode and forward}
\acrodef{CCDF}{complementary cumulative distribution function}
\acrodef{ZF}{zero forcing}
\acrodef{RZF}{regularized zero forcing}
\acrodef{WLLN}{weak law of large number}
\acrodef{SLLN}{strong law of large numbers}
\acrodef{TDD}{Time-division duplex}
\acrodef{EE}{energy efficiency} 
\acrodef{HetNet}{heterogeneous network} 
\acrodef{SCP}{Single Cell Processing}
\acrodef{CBF}{Coordinated Beamforming}

%% file: SupportDocuments/defmetric.tex
\usepackage{color}
\usepackage{dsfont}
\usepackage{bbm}

\def\PST{P_{\mathrm{st}}}
\def\PUT{P_{\mathrm{ut}}}

\def\KS{K_\mathrm s}

\def\LS{\lambda_{\mathrm{s}}}
\def\LU{\lambda_{\mathrm{u}}}

\DeclareMathAlphabet{\mathsf}{OML}{cmbr}{m}{it}

\newtheorem{definition}{\bf Definition}
\newtheorem{theorem}{\bf Theorem}
\newtheorem{lemma}{\bf Lemma}

\newcommand{\bd}{\begin{description}}
\newcommand{\ed}{\end{description}}
\newcommand{\be}{\begin{enumerate}}
\newcommand{\ee}{\end{enumerate}}
\newcommand{\bi}{\begin{itemize}}
\newcommand{\ei}{\end{itemize}}
\newcommand{\bl}{\begin{list}}
\newcommand{\el}{\end{list}}
\newcommand{\bt}{\begin{tabbing}}
\newcommand{\et}{\end{tabbing}}